\documentclass[printnumbers,amsmath,amssymb,twocolumn,superscriptaddress, showkeys,showpacs]{revtex4-1}

\usepackage{graphicx}
\usepackage{natbib}
\usepackage[outdir=./]{epstopdf}

\begin{document}

\title{Localization in one-dimensional chains with L\'{e}vy-type disorder}

\author{Sepideh S. Zakeri}
\email{zakeri@lens.unifi.it}
\affiliation{European Laboratory for Non-linear Spectroscopy (LENS), University of Florence, Via Nello Carrara 1, I-50019 Sesto Fiorentino, Italy}

\author{Stefano Lepri}
\email{stefano.lepri@isc.cnr.it}
\affiliation{Consiglio Nazionale delle Ricerche, Istituto dei Sistemi Complessi, 
via Madonna del Piano 10, I-50019 Sesto Fiorentino, Italy}
\affiliation{Istituto Nazionale di Fisica Nucleare, Sezione di Firenze, 
via G. Sansone 1, I-50019 Sesto Fiorentino, Italy}

\author{Diederik S. Wiersma}
\affiliation{European Laboratory for Non-linear Spectroscopy (LENS), University of Florence, Via Nello Carrara 1, I-50019 Sesto Fiorentino, Italy}
\affiliation {Consiglio Nazionale delle Ricerche, Istituto Nazionale di Ottica, 
Largo Fermi 6, I-50125 Firenze, Italy}
\affiliation {Universit\'a di Firenze, Dipartimento di Fisica e Astronomia, via G. Sansone 1, I-50019 Sesto Fiorentino , Italy}

\begin{abstract}
We study Anderson localization of the classical lattice  
waves in a chain with mass impurities distributed randomly through a power-law relation  $s^{-(1+\alpha)}$ with $ s $ as the distance between two successive impurities and $\alpha>0$.
This model of disorder is long-range correlated and is inspired by the peculiar structure of 
the complex optical systems known as L\'evy glasses. Using theoretical arguments and numerics, 
we show that in the regime in which the average distance between impurities 
is finite with infinite variance, the small-frequency behaviour of the localization length is 
$ \xi_\alpha(\omega) \sim \omega^{-\alpha} $. 
The physical interpretation of this result is that, for small frequencies and long wavelengths,
the waves feel an effective disorder whose fluctuations are scale-dependent. 
Numerical simulations show that 
an initially localized wavepacket attains, at large times, a characteristic 
inverse power-law front with an $ \alpha $-dependent exponent which
can be estimated analytically. 
\end{abstract}


\maketitle

\section{Introduction}
The spatial distribution of disorder plays a key role on transport properties in disordered media, allowing anomalous laws like superdiffusion to arise. This interesting topic has been basis of numerous theoretical and experimental studies in a wide variety of complex systems. Recently, one of the studies reports on realization of some engineered materials named L\'{e}vy glasses in which light rays propagate through an assembly of transparent microspheres embedded in a scattering medium \cite{Barthelemy2008}. If the diameter of the microspheres $ \phi $ is designed to have a power-law distribution $ p(\phi)\sim\phi^{-(\alpha+1)} $, where $ \alpha $ is the so-called L\'{e}vy exponent defining degree of the heterogeneity of the system, light can indeed perform superdiffusion. Although a random walk model is rich and precise to describe light transport through L\'{e}vy glasses and successfully explains the experimental observations \cite{Bertolotti2010}, it cannot address wave properties such as polarization and interference. Therefore, an open question to understand is how interference of waves can affect the propagation and what might be the possible role of Anderson localization \cite{Anderson1958,Sheng2006} in such materials.

Although our main motivation stems from the above described setup, it is worth mentioning some
related studies of systems with L\'{e}vy-type disorder that include transport in quantum wires \cite{Falceto2010}, photonic heterostructures \cite{Fernandez-Marin2012}, 
and disordered electronic systems \cite{Wells2008,Iomin2009}. The purpose of this work is to provide a framework to investigate localization in power-law correlated disordered systems and to illustrate how the localization length is affected by different characteristic features of the system such as frequency, degree of heterogeneity, disorder strength, etc. To address the above issues we 
consider a very simple model, a harmonic chain of coupled oscillators with random 
impurities separated by random distances $s$ having a power-law distribution $ p(s)\propto s^{-(\alpha+1)} $. This model of disorder is clearly inspired by the peculiar structure 
of L\'{e}vy glasses.
As usual, the one-dimensional case allows for a detailed study. In particular, 
the Lyapunov exponent $ \gamma(\omega) $,  the inverse of the frequency-dependent localization length $ \xi(\omega)=\gamma(\omega)^{-1} $ can be computed
straightforwardly. Our main result is that, in the small-frequency regime, the scaling relation $ \gamma(\omega) \propto \omega^ \alpha $ is numerically estimated for the Lyapunov exponent in the range $ 1\leq \alpha\leq 2 $, i.e. when the variance of the 
distances $\langle s^2\rangle$ is infinite. Instead, for $ \alpha>2 $, 
the usual scaling $  \gamma(\omega) \propto \omega^ 2 $, typical of random uncorrelated disorder \cite{Matsuda1970} is recovered.

The model and some theoretical arguments, based on the Hamiltonian map formulation of the transfer method, will be presented in section \ref{theory}.
We then present and discuss the complete numerical steady-state analysis in section \ref{numerics} of this paper. For a more comprehensive understanding of the transport in such disordered media, we investigate and report in section \ref{wavepacket analysis} the time evolution of an initially localized wave packet.
In particular, we consider 
the time and disorder averaged energy profile $ \langle \overline{e_n(t)} \rangle $. Here, we show that in the range $ 1\leq \alpha< 2 $, the asymptotic tails of the energy profile decay with an $ \alpha $-dependent power-law exponent. In contrast, for the range $ \alpha\geq 2 $, depending on the type of the initial excitation, the power-law exponent is independent of $\alpha$. Moreover, we pay close attention to the time evolution of the different moments $ m_\nu (t) $ of the $ \langle \overline{e_n(t)} \rangle $ and discuss the $ \alpha $-dependent properties in the range $ 1\leq \alpha< 2 $. 
Finally, we summarize our results in the concluding section \ref{conclusion}.

\section{Theoretical description}
\label{theory}
\subsection{Model: 1D L\'{e}vy-type disordered lattice}\label{model}

\begin{figure*}
\begin{center}
\includegraphics[width=0.95\textwidth]{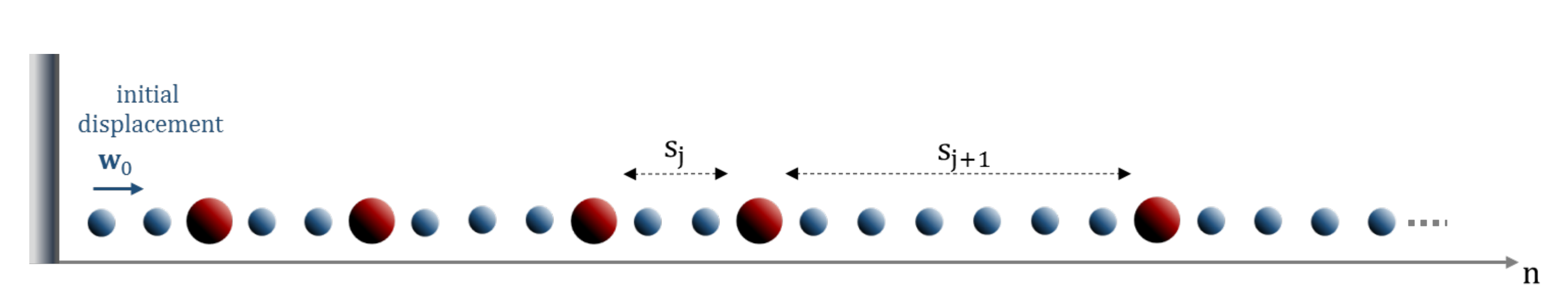}
\caption{(color online) Schematic illustration of the model used to study localization in a one-dimensional lattice with L\'{e}vy-type disorder. Small blue balls represent the background mass of the lattice $ m $ and large red balls are the defects with mass $ M $ distributed on the lattice through a power-law relation $ p(s)\propto s^{-(\alpha+1)} $, where $ s $ is the number of lattice sites between two successive defects and the index $ j=1,...,N_d $. An initial displacement $ \textbf{w}_0 $ is given to the first site of the lattice to initiate iteration of the transfer matrix over the entire lattice.   \label{scheme}}
\end{center}
\end{figure*}

We considered a one-dimensional harmonic disordered lattice of $ n $ sites with masses $ m_n $. 
The governing equations of motion are 
\begin{equation}
m_n\ddot u_n = k(u_{n+1}-u_n) + k(u_{n-1}- u_n),
\label{eqmotion1}
\end{equation} 
where $ k $  is spring constant ($ k\neq 0 $), $ u_n $ is displacement of the $ n $th site of the lattice from its equilibrium position and $  n=1,2,..., N $. 

In the present work, we consider dichotomic disorder whereby 
$ m_n $ assumes two possible values, either $ m $ as the background mass of the lattice or $ M $ to represent mass of the defects. Illustrated in Fig. \ref{scheme}, disordered lattices of length $ N $ were constructed by arrangement of $ N_d $ defects through a power-law distribution $ p(s)\propto s^{-(\alpha+1)} $, where $ s $ was the random number of lattice sites with mass $ m $ between two successive defects. As a direct consequence of such choice of $ p(s) $, the mass of the entire lattice was the parameter with $ \alpha $-dependent statistical properties. Obviously, its average density can be written as
\begin{eqnarray}
&& \langle m \rangle=\rho M+(1-\rho)m,
\end{eqnarray} where $ \rho=\frac{N_d}{N} $ is the fraction of defects on the lattice. In the range $ 1\leq \alpha <2 $, $ \langle m \rangle $ is finite. On the other hand, for $ \alpha <1$, 
$\rho$ vanishes in the thermodynamic limit $N\to \infty$ since the average distance between
consecutive defects diverges in this regime. In this paper, we mainly focus on the case
$ \alpha > 1 $ (in which $ \langle s \rangle $ is finite) 
and will comment only briefly on the case $ \alpha < 1$ which is somehow more peculiar.

Before proceeding, we recall that random walks on such class of L\'{e}vy structures have 
been thoroughly studied in a series of recent papers  \cite{Barkai2000,Beenakker2009,Barthelemy2010,Burioni2010,Buonsante2011,Bernabo2014} 
as a minimal model that includes quenched disorder and
anomalous diffusion. Depending on the value of $ \alpha $, fundamentally different regimes of transport are achieved: indeed walkers superdiffuse for $\alpha<1$. Several distinguished 
features of the quenched nature of disorder like the importance of initial 
conditions and the consequences on higher-order statistics of the diffusive 
process are discussed in the above mentioned papers. It can be thus envisaged that also 
localization properties may display unusual features.

\subsection{Statistical properties of the disorder}

\begin{figure}
\begin{center}
\includegraphics[width=0.4\textwidth]{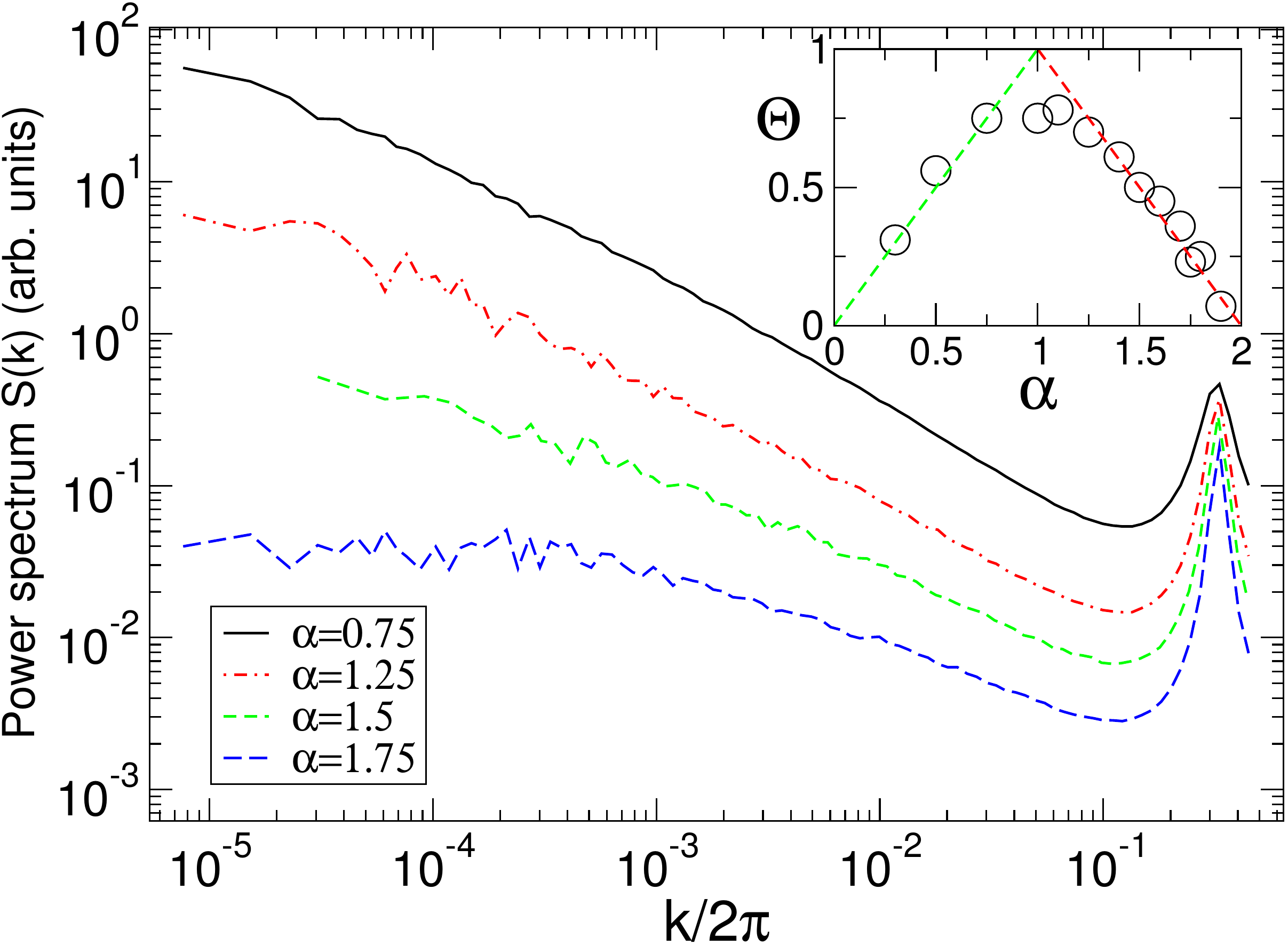}
\caption{(color online) Power spectra, Eq.~(\ref{spectrum}), for different values of the exponent
$\alpha$; each spectrum is obtained for sequences of length $2^{17}$ over an ensemble of 
$10^4$ realizations. The inset shows the dependence of the exponent of the small-frequency 
singularity $|k|^{-\Theta}$ on $\alpha$.   
\label{f:sk}}
\end{center}
\end{figure}

Using the algorithm described in \cite{Devroye1986}, positive integer power-law random numbers were generated to construct disordered lattices. For a preliminary statistical characterization, we 
computed the ensemble-averaged power spectrum:
\begin{equation}
S(k)= \frac{1}{N} \left\langle \big|\sum_{n=1}^N m_n \exp(-ikn)\big|^2\right\rangle, 
\label{spectrum}
\end{equation}
where the average is over different realization of the disorder. The low-wavenumber
behavior of $S$ provides information on the large-scale decay law of the 
disorder correlation function which will be necessary for the subsequent analysis.
Fig. \ref{f:sk} indicates that, for $\alpha < 2$, 
$S$ has a power-law singularity $|k|^{-\Theta}$ for small $k$. 
This implies that the disorder correlation is indeed decaying as an inverse power of 
the relative distance with an exponent $1-\Theta$. The exponent will be important in
the subsequent analysis: fitting of the numerical data suggests the following relation with the L\'{e}vy
parameter $\alpha$ (see the Inset of Fig. \ref{f:sk})
\begin{equation}
\Theta(\alpha) = \begin{cases} 
\alpha  &\qquad \mbox{if} \quad 0<\alpha<1 \\
2-\alpha  &\qquad \mbox{if} \quad 1<\alpha<2 \end{cases}.
\label{eq:beta}
\end{equation}
A theoretical justification of the above relation can be given based on the 
similarity of this process with the correlation of a L\'{e}vy walk as discussed  
by Geisel in Ref.~\cite{Geisel1995}.
Note that in all the examined cases $\Theta<1$, so the power spectrum is integrable 
and the associated process can be considered as stationary for any $\alpha$.

For what concerns the dependence on the 
lattice size (data not shown), by comparing data for different lengths, we found that for $\alpha>1$ the spectra are $N$ independent, while for $\alpha<1$ the spectra decrease with increasing $N$ 
roughly like $N^{\alpha-1}$. This is because the density of defects vanishes in the thermodynamic
limit and almost-ordered realizations dominate the statistical averages.  

To conclude this subsection, let us mention that a similar model has 
been studied in Ref.~\cite{Moura2003} (see also the related work 
\cite{Costa2011}). The main differences with our work is that
the sequence of masses is generated as a trace of a fractional 
Brownian motion that, by construction, has also a a power-law
singularity at small wavenumber. It was there shown that in the 
non-stationary case ($\Theta>1$ in our notation), a mobility 
edge can exist at a finite frequency value.  

\subsection{Transfer matrix approach}
\label{tmat}

As it is well known, the localization properties can be studied by the transfer matrix method 
\cite{Crisanti1993}.
Assuming that $ u_n $ oscillates harmonically in time at an angular frequency $ \omega $, so that $u_n =v_n e^{i\omega t}$, Eq. (\ref{eqmotion1}) can be written as an eigenvalue problem
\begin{equation}
-m_n \omega^2 v_n = k(v_{n+1}-v_n) + k(v_{n-1}- v_n),
\label{eqmotion2}
\end{equation} 

that, upon defining 
\[ \boldsymbol{w_n} \equiv \begin{pmatrix} 
  v_{n} \\ 
  v_{n-1}
\end{pmatrix}
;\quad 
\boldsymbol{T_n}\equiv \begin{pmatrix}
   2-m_n\omega^2/k  &&& -1 \\
  1  &&&  0
\end{pmatrix}, 
 \] for $  n=2,3,..., N-1 $ can be recast in the form 
\begin{equation}
 \boldsymbol{w_{n+1}}=\boldsymbol{T_n}\boldsymbol{w_n},
\end{equation} 
where $ T_n $ is a $ 2\times 2 $ transfer matrix of the $ n $th site on the lattice. By iteration of the transfer matrix over the entire lattice and applying the appropriate boundary conditions of the system, solutions of Eq. (\ref{eqmotion2}) as a function of frequency $ \omega $ can be obtained. 

The central quantity to be computed is the Lyapunov exponent $\gamma(\omega)$ which gives 
the inverse of the localization length $\xi(\omega)$.
It is convenient to define a new parameter $ R_n=v_n/v_{n-1} $. Inserting $ R_n $ into Eq. (\ref{eqmotion2}), one obtains the following recursive relation \cite{Derrida1984}
\begin{equation}
R_{n+1}=2- \frac{m_n\omega^2}{k}-\frac{1}{R_n}.
\label{eqrecursive}
\end{equation}
Eq. (\ref{eqrecursive}) can be interpreted as a "discrete time" stochastic equation. The mass $m_n$ plays the role of a noise source (with bias) whose strength is gauged by the frequency $\omega$. In the present model, Lyapunov exponent $\gamma(\omega)$ as the inverse of localization length can be computed by
\begin{equation}
    \xi(\omega)^{-1}=\gamma(\omega) = \langle \ln R_n \rangle. 
\label{eq:lyap}
\end{equation}
Also the integrated density of states $I(\omega)$ follows from node counting
arguments, i.e. $I(\omega) = f$, where $f$ is the fraction of  negative $R_n$
values.

For deeper analysis of the systems we are interested in, let us rewrite Eq. (\ref{eqmotion2}) in a different notation, introducing the new variables $q_n=v_n$ and $p_n=q_n-q_{n-1}$, the transfer map can be reformulated as follows. Letting $\omega^2m_n = \Omega^2(1+\mu_n)$, $\Omega^2=\omega^2 \langle m \rangle$  and $\mu_n=m_n/ \langle m \rangle-1$ is a zero-average random variable and $\langle \mu^2\rangle$ is finite at least for $\alpha>1$. Plugging $ q_n $ and $ p_n $ into Eq. (\ref{eqmotion2}) and setting $ k=1 $, one finds the following two-dimensional map \cite{Izrailev1998}
\begin{eqnarray}
&& p_{n+1} = p_n - \Omega^2(1+\mu_n)q_n \label{map-p}\\
&& q_{n+1} = q_n + p_{n+1}.
\label{map-q}
\end{eqnarray} 

The following transformation relations can be introduced by using the canonical variables $ (r_n, \theta_n) $ where $ r_n $ and $ \theta_n $ are amplitude and phase of the eigenvector at site $ n $, respectively:
\begin{equation}
p_n = \sqrt{2\Omega} \, r_n \sin \theta_n;\quad
q_n = \sqrt{\frac{2}{\Omega}} \, r_n \cos \theta_n.
\label{var-q}
\end{equation}
Substituting Eqs. (\ref{var-q}) back into Eqs. (\ref{map-p}) and (\ref{map-q}) and neglecting terms of the order $\Omega^2$ and higher, one then obtains the following map
\begin{eqnarray}
&&\left(\frac{r_{n+1}}{r_n}\right)^2= 1-\mu_n\Omega \sin 2\theta_n +O(\Omega^2)\\
&& \tan  \theta_{n+1} =  \tan(\theta_{n} -\Omega ) -\Omega \mu_n. \label{phase1}
\end{eqnarray}
Eq. (\ref{phase1}) describes evolution of the phase as it is perturbed by disorder and can be approximated as \cite{Izrailev2012}
\begin{equation}
\theta_{n+1} =  \theta_{n} -\Omega + \Omega \mu_n \sin^2 \theta_n. \label{phase2}
\end{equation}
Long-range correlations of $\mu_n$ lead to sub-diffusive phase fluctuations. Within the same approximation, the Lyapunov exponent is obtained by expanding the logarithm
\begin{equation}
\begin{split}
\gamma = \frac 12\langle \log \left(\frac{r_{n+1}}{r_n}\right) \rangle \approx
-\frac 12 \Omega \langle\mu_n \sin 2\theta_n\rangle \\
+ \frac 14 
\Omega^2 \langle\mu_n^2 \sin ^2 2\theta_n\rangle +\ldots
\end{split}
\label{eq:gammaapp}
 \end{equation}
 
The main issue is to estimate the correlators in Eq. (\ref{eq:gammaapp}).
Following the same path as in ref. \cite{Izrailev2012} (see in particular 
section (5.2)), it can be shown that 
\begin{eqnarray}
&& \langle\mu_n \sin 2\theta_n\rangle = 2\Omega \sum_{m=1}^\infty 
\langle\mu_n \mu_{n-m} \rangle \cos (2\Omega m)\\
&& \langle\mu_n^2 \sin ^2 2\theta_n\rangle \approx \frac12  \langle\mu_n^2\rangle.
\end{eqnarray}
We should get something like
\begin{equation}
\gamma = \frac18 \Omega^2\langle\mu_n^2 \rangle \left( 1 +2\sum_{m=1}^\infty 
\frac{\langle\mu_n \mu_{n-m} \rangle}{\langle\mu_n^2 \rangle} \cos (2\Omega m) \right).
\label{eq:agamma}
\end{equation}
Note that this formula agrees with the long-wavelength limit of Eq.(22) in \cite{Herrera-Gonzalez2010} which was obtained for weak disorder 
(where $\Omega\approx k$).
The sum is basically the Fourier transform of the correlation function, i.e.
the power spectrum of the sequence of masses studied above. 
We thus expect three cases:
\begin{itemize}
\item For short-range correlations as in the case $\alpha>2$, the second term in Eq.
(\ref{eq:agamma}) is finite and independent of the frequency. So it just provides 
a proportionality constant. The Lyapunov exponent is thus expected to scale, up to some constant prefactor as in the standard uncorrelated case \cite{Matsuda1970}, 
\begin{equation}
  \gamma(\omega) \propto \omega^2\frac{ \langle {m}^2\rangle - \langle m \rangle ^2}{8 \langle m \rangle}.
%  ,\qquad {\rm for} \qquad \omega \longrightarrow 0
\label{eq:analoc}
\end{equation} 

\item For long-range correlations and in the case $1<\alpha<2$, the 
fluctuation spectrum diverges at small frequencies so the second term
in Eq. (\ref{eq:agamma}) dominates so that
 \begin{equation}
\gamma \sim   \Omega^2 S(2\Omega).
\end{equation}
Using the result Eq.~(\ref{eq:beta}) that the spectrum goes as $k^{\alpha-2}$
and recalling the definition of $\Omega$ this 
means that the Lyapunov exponent should be proportional to $\omega^\alpha$. 

\item Furthermore, if we assume that the same argument holds also for $\alpha<1$,
we may conclude that $\gamma \sim \omega^{2-\alpha}/N^{1-\alpha}$, which means that $\gamma$ vanishes for $N\to\infty$. This argument may not be entirely correct since in this case, $\langle\mu_n^2 \rangle$ also
vanishes.
\end{itemize}

To conclude this section, we mention that John and Stephen \cite{John1983} studied a related model of a classical wave in the 
continuum limit where the mass fluctuations $\mu$ were quenched random variables 
but power-law correlated 
in space $\langle \mu(x)\mu(x')\rangle \sim (x-x')^{-2m}$. According to their Eq. (7.5), in 1D
the Lyapunov exponent should be
\begin{equation}
 \gamma (\omega) \sim
 \begin{cases} \omega^2 &\mbox{if } m>1/2 \\
\omega^{\frac{1}{1-m}} & \mbox{if } m<1/2 
\end{cases}. 
\end{equation}
Although the models are pretty similar at large scales, our result is different 
from this last estimate. The origin of the discrepancy is not clear at this stage, however 
it is true that the approach in ref. \cite{John1983} relies on some approximate
self-consistent calculations while the calculations reported here are somehow exact.

\section{Numerical results}
\label{numerics}

In this section, numerical results of the model described in the previous section are presented.  We have carefully studied the dependence of the Lyapunov exponent on different characteristic parameters of the systems. Data analysis was performed mainly for systems with $1\leq\alpha< 2 $, however, some computations were done for systems with $ \alpha\geq 2 $ to allow comparison of the behaviours in two different transport regimes. Throughout the rest of the paper, the spring constant was assumed to be $ k=1 $ and, unless otherwise stated, the 
mass ratio was fixed to $ M/m=3 $.

\subsection{Lyapunov exponent}\label{Lyapunov analysis}
\begin{figure}
\begin{center}
\includegraphics[width=0.45\textwidth]{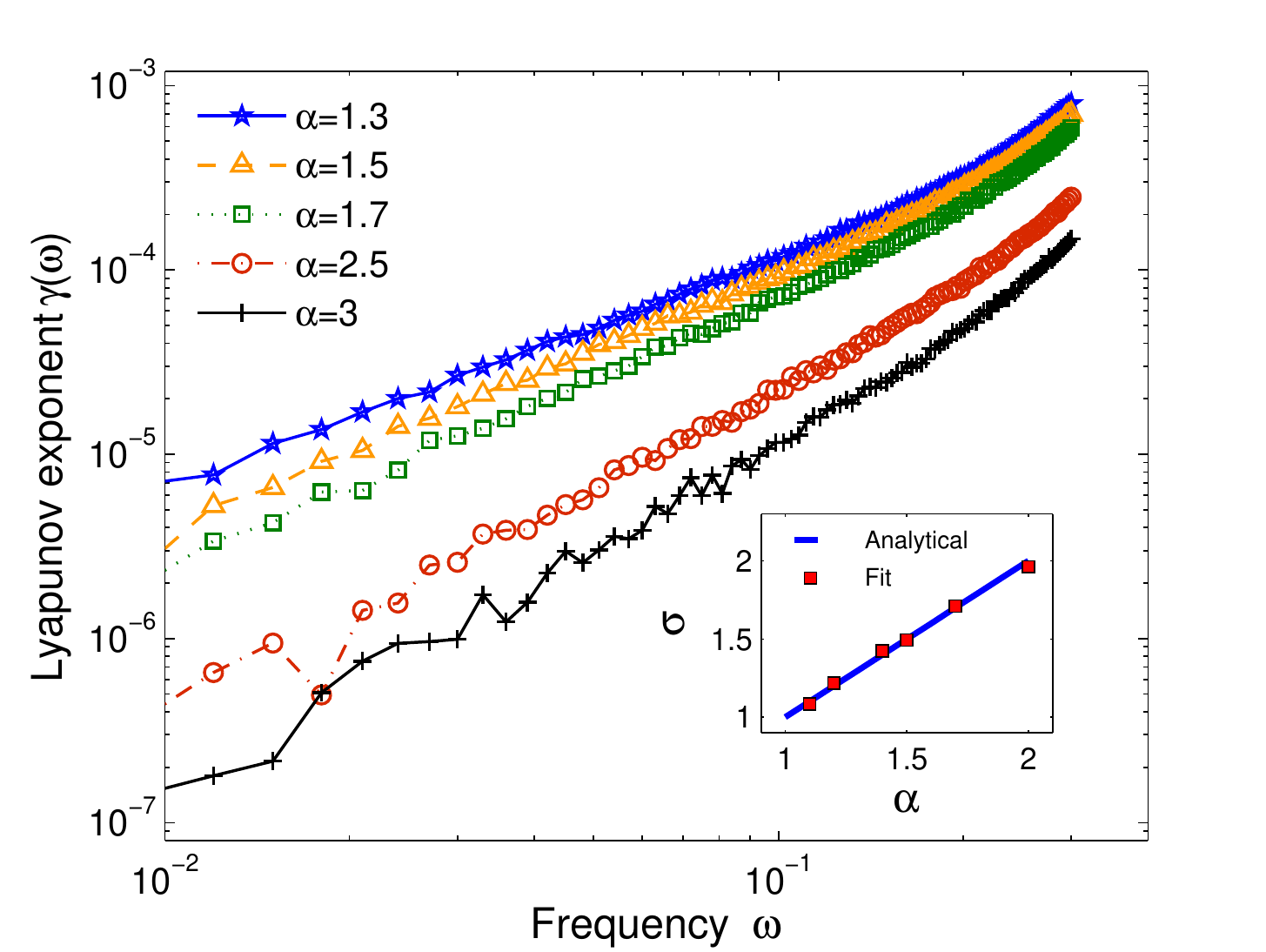}
\caption{(color online) Variations of the Lyapunov exponent $ \gamma(\omega) $ versus $ \omega $ for systems with different $ \alpha $ parameter, setting $ N_d=5\times 10^6 $, $ M/m=3 $ and $ k=1 $. Lyapunov exponents were normalized after $ N_{it}=5\times 10^4 $ iterations of the transfer matrix. Frequency $ \omega $ was varied from $ 0.01 $ to $ 0.3 $. The inset shows the scaling exponent $ \sigma $ (of the frequency-dependent Lyapunov exponent $ \gamma(\omega)\approx \omega^\sigma $) versus $ \alpha $. Numerical results for the scaling are obtained by power-law fits on the data set for Lyapunov exponent and an excellent agreement is obtained with the theoretical predictions $ \gamma(\omega)\approx \omega^\alpha $ in the low-frequency regime for the range $ 1<\alpha<2 $. \label{Lyapomega}}
\end{center}
\end{figure}

According to the definition described in Eq. (\ref{eq:lyap}), the Lyapunov
exponent is the exponential growth or decay rate of a vector in the
limit $ N\rightarrow \infty $. 

Due to the long-range spatial correlations 
(especially in the regime $ 1\leq\alpha<2 $), the convergence of the 
numerical value of $\gamma$ could vary from
a single realization to another and usually requires a long series of 
iterations. In order to ensure that our obtained results
were independent of the disorder realization (i.e. that self-averaging holds), 
a series of lattices with different
number of defects $ N_d $ were studied and compared. Results led us to the
conclusion that for $ N_d \geq 2\times 10^6 $, fluctuations of $ \gamma(\omega)
$ versus $ N_d $ are relatively small (better than $\simeq \pm 13 \% $) in the systems with
different $ \alpha $ parameters. Clearly, as $ \alpha $ approaches to $ 2 $,
even smaller values of $ N_d $ can yield a fast convergence  $ \gamma(\omega)
$. However, to improve stability of our results even further, $ N_d= 5\times
10^6 $ defects were used to construct our desired disordered systems. We pursue
this number consistently in the rest of this paper for the analysis of very
large lattices (asymptotic limit).
In all the studied cases, same initial displacement $ \bf w_0 $ must
be assigned to start iteration of the transfer matrix. Note that Lyapunov exponent
has this intrinsic property to be independent of $ \bf w_0 $ and our
numerical results were in excellent agreement with that. 

Since we mostly focus on the behaviour of the Lyapunov exponent in the small
frequency regime, we firstly checked that in all the presented cases 
the integrated density of states is linearly changing with the frequency, $I(\omega) \sim
\omega$. The physical interpretation of this fact is simply that the 
spectrum is effectively equivalent to an ordered chain with a renormalized
wave speed. 

In Fig. \ref{Lyapomega}, variations of the Lyapunov exponent $ \gamma(\omega) $
versus frequency $ \omega $ for systems with different $ \alpha $ parameter are
shown in doubly logarithmic scale. In each system, as the frequency was increased from $
0.01 $ to $ 0.3 $, greater values were obtained for Lyapunov exponent and accordingly, localization length was decreased. As illustrated in the inset of Fig. \ref{Lyapomega}, the scaling relation predicted in the previous section is in excellent agreement with the data obtained by power-law fitting on the small-frequency region of the curves.
In some cases, we also checked that the fit that includes the leading 
order correction $\omega^2$ follows the data in the whole range,
giving further support to the theoretical arguments above.

\begin{figure}
\begin{center}
\includegraphics[width=0.45\textwidth]{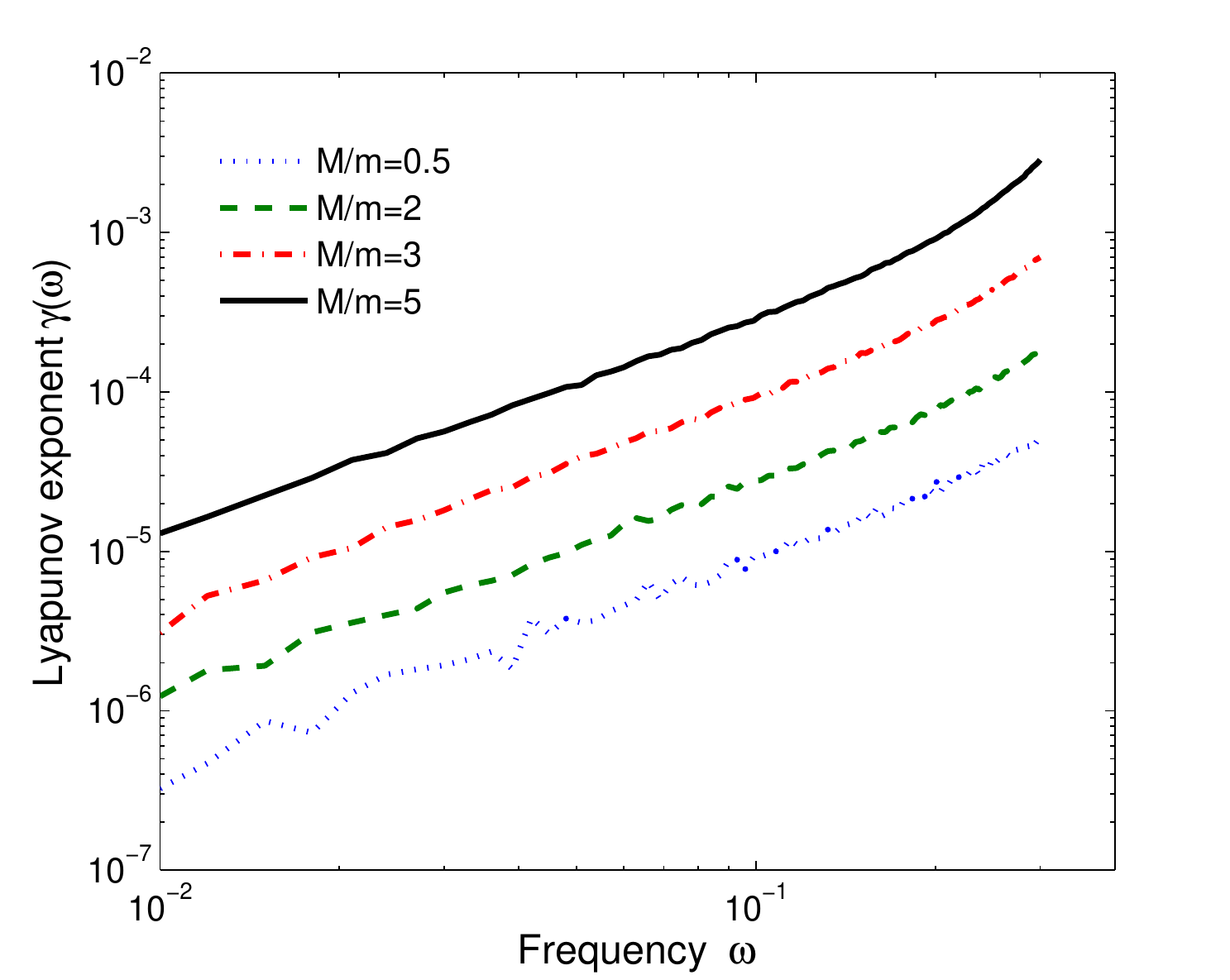}
\caption{(color online) Comparison of the Lyapunov exponent $ \gamma(\omega) $ versus $ \omega $ for systems with $ \alpha=1.5 $ and $ N=14648365 $ but different mass ratios $ M/m $.  Other parameters such as $ N_d $, $ N_{it} $ and the frequency range are the same as those used to generate Fig. \ref{Lyapomega}. \label{mass}}
\end{center}
\end{figure}

\begin{figure}
\includegraphics[width=0.45\textwidth]{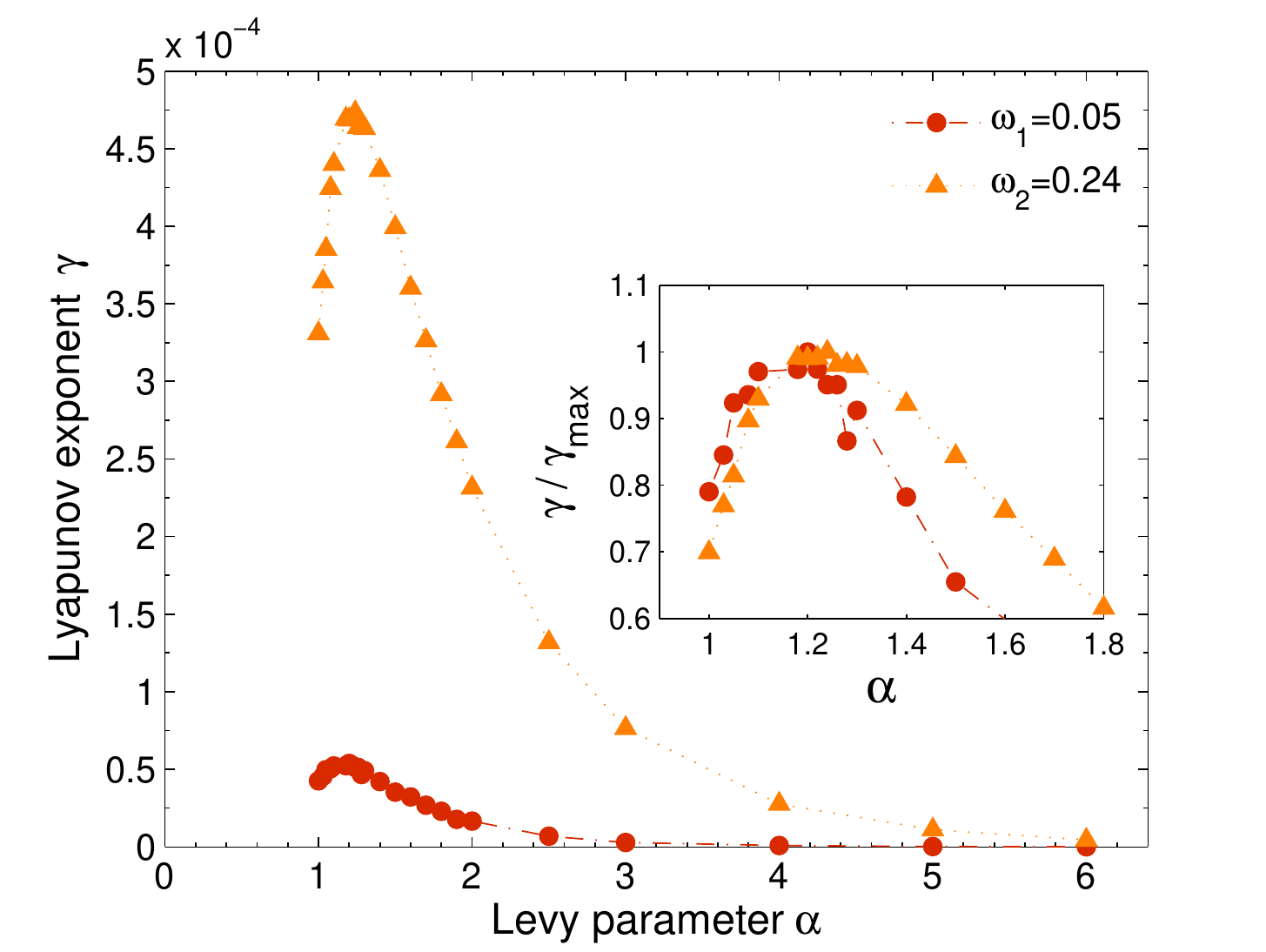}
\caption{(color online) Dependence of the Lyapunov exponent on the L\'{e}vy exponent $ \alpha $ at two fixed frequencies $ \omega_1=0.05 $ (red circles) and $ \omega_2=0.24 $ (orange triangles). The inset shows that by increase of the frequency, peak of the Lyapunov exponent is shifted to right. For better illustration, two curves are normalized by their maxima. \label{Lyapalpha}}
\end{figure}

Another interesting physical effect that should be assessed is the dependence
of the Lyapunov exponent on disorder strength or equivalently mass ratio $ M/m $
in our model. In an intuitive picture, one expects that higher mass ratio leads
to stronger disorder, shorter localization length and thereby greater Lyapunov
exponents. In addition to that, the scaling of $ \gamma(\omega)\propto \omega^\sigma
$ intrinsically depends on transport properties of the system and hence should 
be independent of the mass ratio. In other words, the mass ratio
should at most change the prefactor in the above scaling relation. To confirm
this prediction, four different mass ratios $ M/m=0.5, 2, 3, 5 $ were considered
in a system with $ \alpha=1.5 $ and length $ N=14648365 $ composed of $
N_d=5\times 10^6 $ defects. As shown in Fig. \ref{mass}, scaling of the Lyapunov exponent with 
frequency $ \omega $ (slope of the curves) is independent of the mass ratios.
However, larger values of $ M/m $ yield an increase in the Lyapunov exponent by orders of magnitude, indicating that the prefactor changes rapidly with  $ M/m $.

Fig. \ref{Lyapalpha} shows and compares dependence of the Lyapunov exponent on the L\'{e}vy exponent $ \alpha $ at two fixed frequencies $ \omega_1=0.05 $ and $ \omega_2=0.24 $. 
Upon decreasing $\alpha$, $\gamma$ was initially increased. This is in agreement with the intuition that the localization is enhanced by increasing fluctuations of the distances between defects. However, by approaching $ \alpha=1 $, $\gamma$ started decreasing and a peak was observed in both curves. We can therefore conclude that in the range $ 1.2 \leq \alpha \leq 1.3$, a transition occurs in the behaviour of the Lyapunov exponent.  The value of $ \alpha $, at which the transition happens, depends slightly on the frequency $ \omega $. According to our numerical data, the transition 
appeared at $ \alpha=1.2 $ for $ \omega_1 $ and at $ \alpha=1.24 $ for $ \omega_2 $ 
(see the inset in Fig. \ref{Lyapalpha}, showing the Lyapunov exponents normalized by their maximums).
Moreover, no appreciable dependence on the lattice length is observed in this regime. Although it seems plausible that $\gamma$ vanishes below $\alpha=1$, the origin of the peak is not clear. A possibility is that sample-to-sample fluctuations become 
so important that averaging over a much larger ensemble of lattices would be required.

\subsection{Eigenfunctions}

Moreover, it is interesting to look at eigenfunctions of a L\'{e}vy-type disordered system. In general, spatial part of the frequency dependent solutions of Eq. (\ref{eqmotion1}) can be obtained by solving
\begin{equation}
(\boldsymbol{K}+ \omega^2 \boldsymbol{\varLambda})\textbf{u}=0, \label{eigenvalueproblem}
\end{equation} where $ \boldsymbol{\varLambda} $ is the diagonalized mass matrix $  \boldsymbol{\varLambda}=$diag$(m_n) $ and $ \boldsymbol{K} $ is the matrix of spring constants expressed by
\[ \boldsymbol{K}=\begin{pmatrix}
   -2k  && k     && 0   && 0     && \hdots\\
   k    && -2k   && k   && 0     &&  \hdots\\
   0    && k     && -2k && k     &&  \hdots\\
\vdots  &&       &&     &&\ddots &&  \\
   0    && \hdots&&  0  && k     && -2k
\end{pmatrix}_{N\times N}  \]

\begin{figure}
\includegraphics[width=0.45\textwidth]{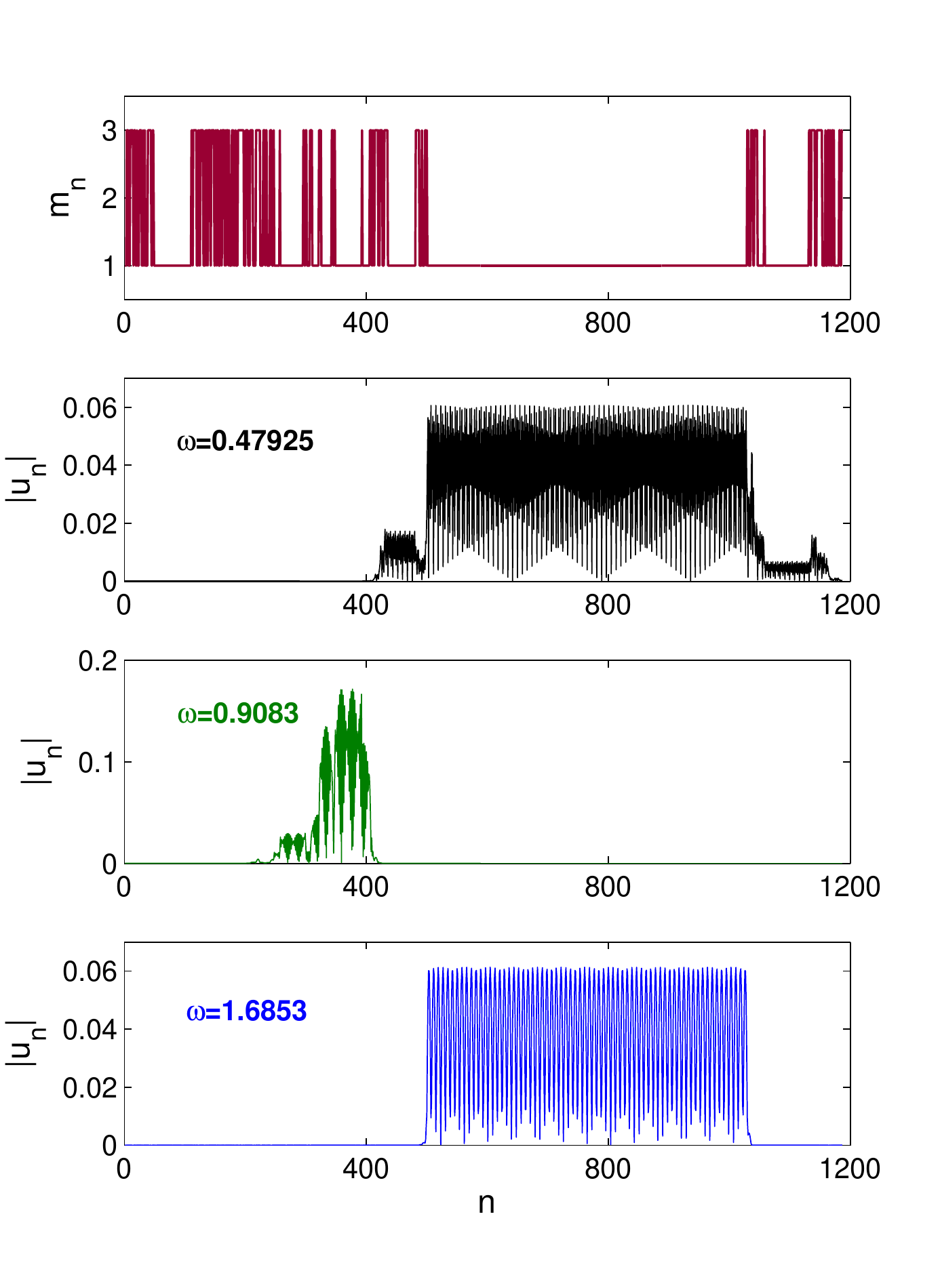}
\caption{(color online) Top panel shows distribution of the mass on a system with $ \alpha=1.1 $ and length $ N=1187 $ composed of $ N_d= 200$ defects (red curve). Other panels (from top to bottom) show eigenfunctions of the system at $ \omega=0.47925 $ (black curve), $ \omega=0.9083 $ (green curve) and $ \omega=1.6853 $ (blue curve). \label{eigenfunction}}
\end{figure}

Stationary eigenfunctions of disordered systems can be obtained by numerically solving Eq. (\ref{eigenvalueproblem}). For a system with $ \alpha=1.1 $ and length $ N=1187 $ consisted of $ N_d= 200$ defects, eigenfunctions $ u_n $  at three different frequencies are presented in Fig. \ref{eigenfunction}. Top panel shows distribution of mass $ m_n $ on the entire system which is a step-function of $ m_n=1 $ or $ 3 $. Eigenfunctions of the systems at three different frequencies are depicted in the other panels.  According to our results, at $ \omega= 0.47925$ and $ \omega=1.6853 $, eigenfunctions are almost extended in a long space between two successive defects. At $ \omega=0.9083 $, however, eigenfunctions are localized in a small region inside the system. Note that because of the computational difficulties to diagonalize the system matrix, for this analysis we could only study finite systems with limited number of defects.

\subsection{Phase dynamics}

\begin{figure}
\includegraphics[width=0.45\textwidth]{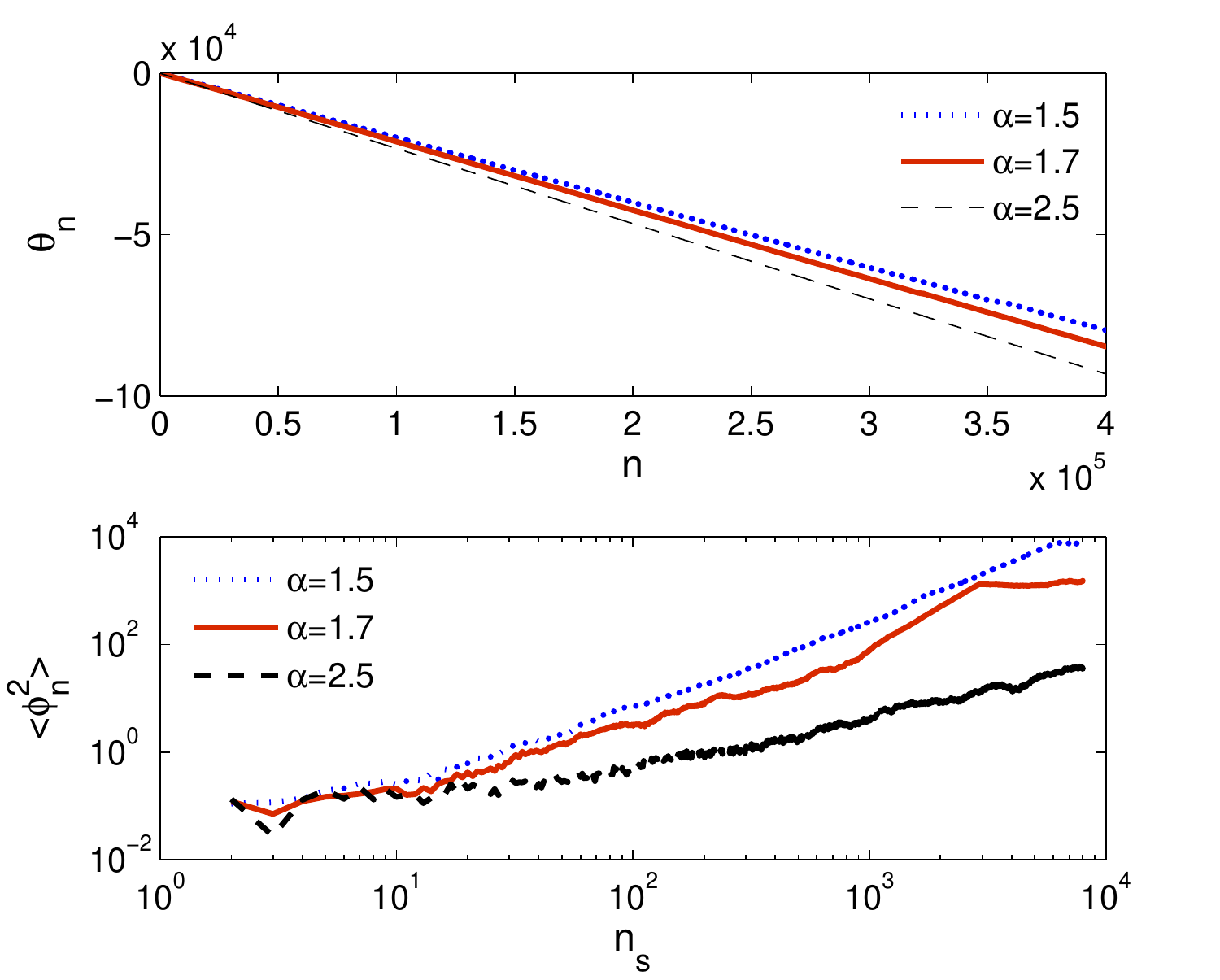}
\caption{(color online) Top panel shows variations of the unwrapped phase $ \theta_n $ versus $ n $ for three different systems with $ \alpha=1.5,1.7,2.5 $ and length $ N= 4\times 10^5$ at a fixed frequency $ \omega=0.1 $. In the bottom panel, variations of $ \langle \phi_n^2 \rangle $ over a sub-lattice of length $ n_s= 8\times 10^3 $ are presented and compared for the studied systems. Note that $ \Omega _{1.5}=0.1297 $, $ \Omega _{1.7}=0.1329 $, $ \Omega _{2.5}=0.1383 $, respectively. \label{phasefig}}
\end{figure}

Next we present our numerical results for the spatial evolution of the phase that, using Eq. (\ref{var-q}) and the definition $ p_n=q_n-q_{n-1} $, can be computed 
during the iteration of the transfer matrix by the following relation
\begin{equation}
\theta_n =tg^{-1}\left( \dfrac{q_n-q_{n-1}}{\Omega q_n}\right). \label{teta}
\end{equation} 
The computed phase has to be unwrapped to the real axis. Top panel in Fig. \ref{phasefig} illustrates variations of $ \theta_n $ in systems of equal length $ N= 4\times 10^5$ and different L\'{e}vy exponents $ \alpha=1.5,1.7,2.5 $ at $ \omega=0.1 $.

For the statistical analysis, the average drift rate $a$ of the phase ($ a \approx \Omega $) was 
numerically computed and removed by defining $ \phi_n=\theta_n-an $. 
In order to study evolution of $ \phi_n $ over the entire system, the lattice was divided into $ 50 $ sub-lattices of length $ n_s=8\times 10^3 $. For each sub-lattice, the quantity $ (\phi_{n}-\phi_1)^2 $, where $ n=1,..,n_s $ was computed. Averaged over all the sub-lattices, evolution of the rms-phase $ \langle \phi_n^2 \rangle $ was consequently obtained as shown in the bottom panel of Fig. \ref{phasefig}. According to the data, growth of $ \langle \phi_n^2 \rangle $ 
is faster in systems with smaller values of $ \alpha $. This can be explained by the fact that disorder fluctuations become stronger which is in agreement with the superdiffusive 
behaviour of the corresponding random-walk problem \cite{Burioni2010}. 

\section{Evolution of wavepackets}\label{wavepacket analysis}

In this section, we investigate the consequences of the above results on 
the spreading of an initially localized perturbation on an infinite lattice.
Although every single eigenstate is exponentially localized, the 
wave propagation can be non trivial as low-frequency states  
form a continuum  with arbitrarily large localization lengths.

The starting point of this analysis is a Hamiltonian formulation of the system:
\begin{equation}
H=\sum_n [\frac{P_n^2}{2m_n}+\frac{1}{2}k(u_{n+1}-u_n)^2],
\end{equation} where $ u_n $ and $ P_n $ are displacement and momentum of the mass $ m_n $, respectively. From the Hamiltonian, time evolution of any excitation of interest can be derived. Evidently, local energy at each site of the lattice at time $ t $ can be written as:

\begin{eqnarray}
&& e_n(t)=e_n^{kin}(t)+e_n^{pot}(t) \label{H1}\\
&& e_n^{kin}(t)=\frac{1}{2}m_n\dot{u}_n(t)^2 \label{H2}\\
&& e_n^{pot}(t)=\frac{1}{2}k[u_n(t)-u_{n-1}(t)]u_n(t)- \nonumber \\
&&\frac{1}{2}k[u_{n+1}(t)-u_n(t)]u_n(t), \label{H3}
\end{eqnarray} where $ e_n^{kin}(t) $ and $ e_n^{pot}(t) $ represent kinetic and potential energies, respectively. Together with the equation of motion Eq. (\ref{eqmotion2}), these relations constitute the set of governing equations of one-dimensional linear discrete systems. 
\begin{figure}
\resizebox{0.9\columnwidth}{!}{
\begin{tabular}{c} 
\includegraphics[width=\textwidth]{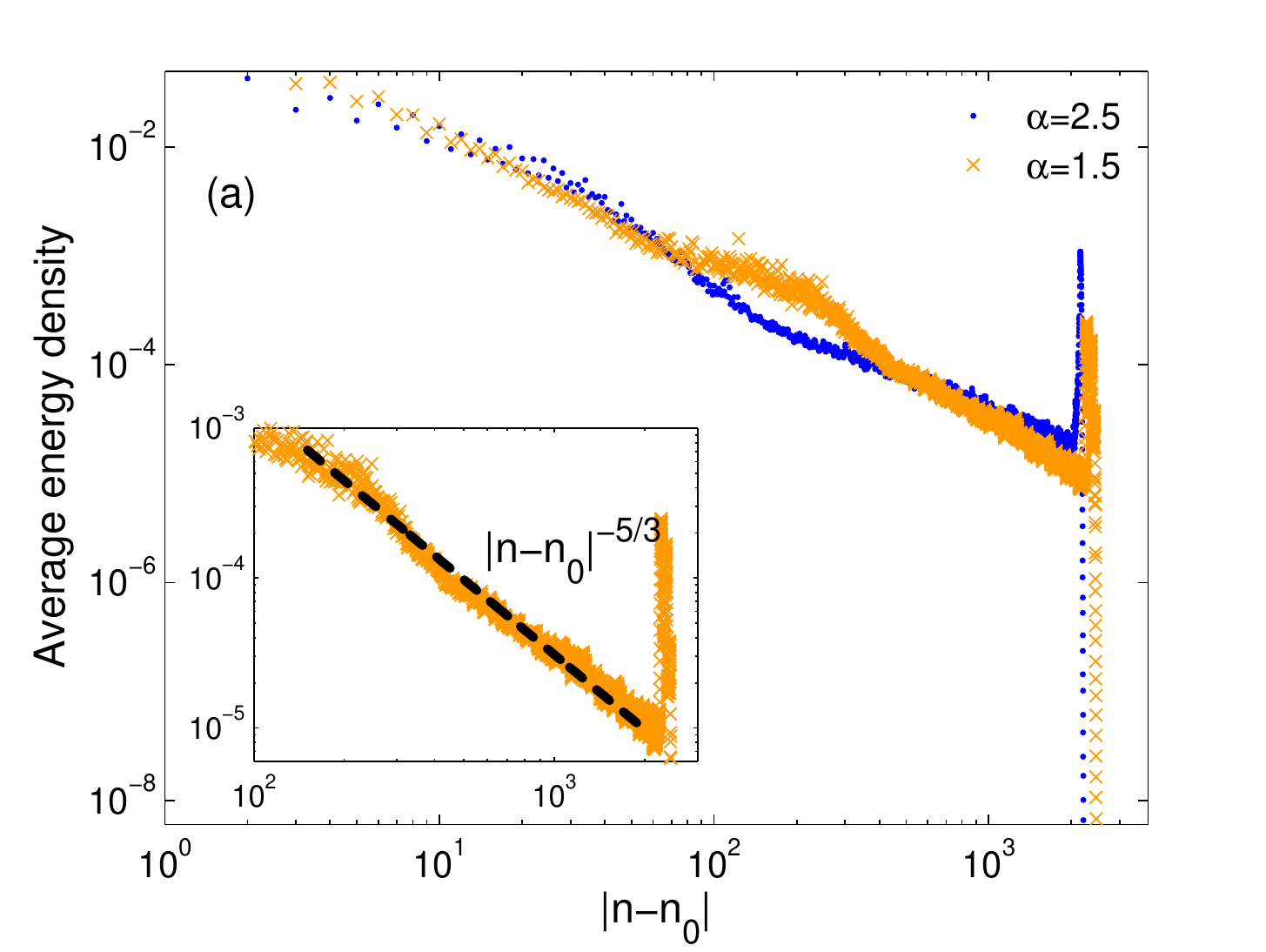} \\ 
\includegraphics[width=0.95\textwidth]{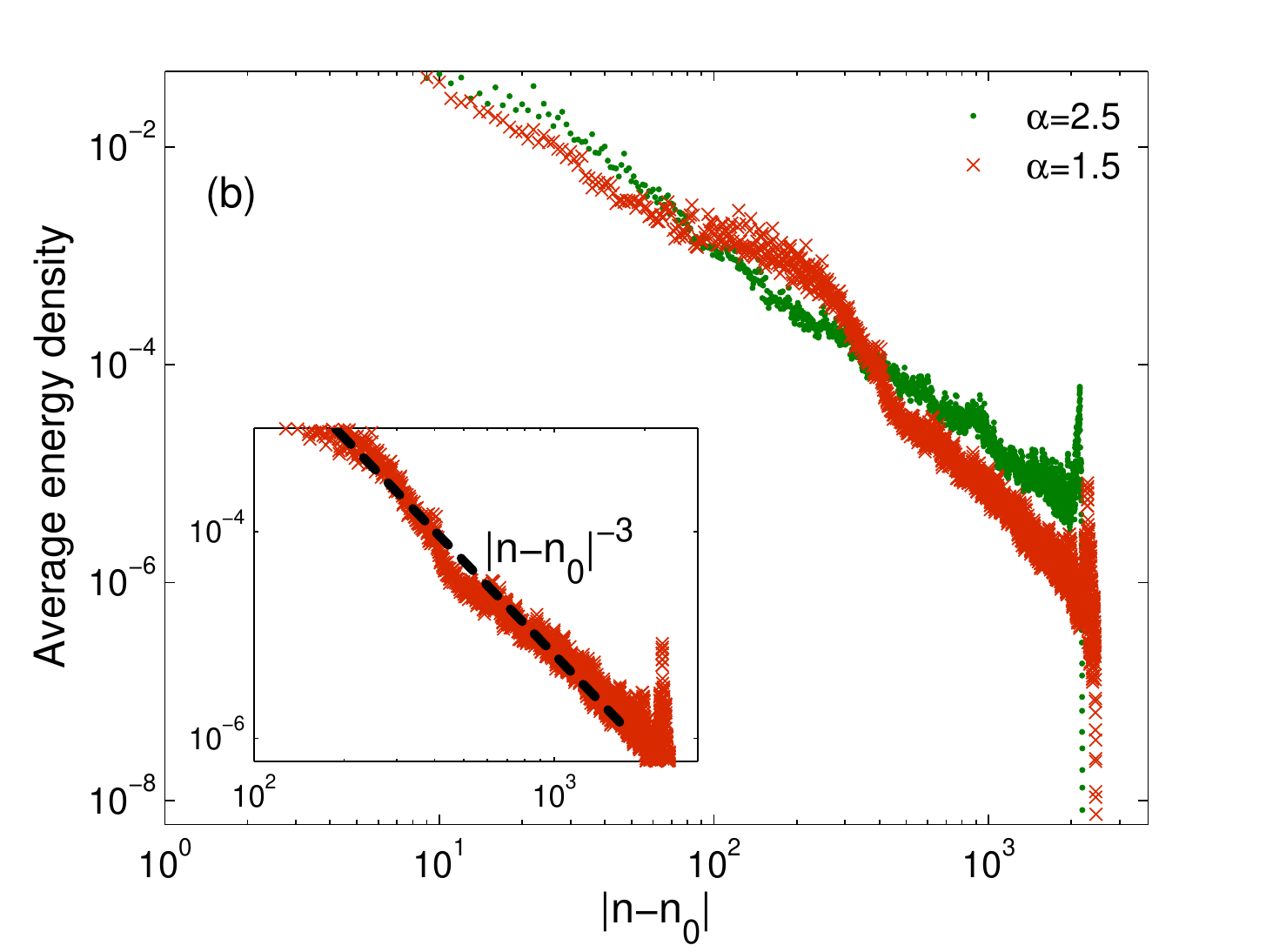}
\end{tabular}}
\label{diffusive fit}
\caption{(color online) Spreading of the energy density after $ t= 3 \times 10^3 $ averaged over $ 2 \times 10^3 $ realizations. Results are comparison for two systems with equal lengths $N=8 \times 10^3  $ but different L\'{e}vy exponents $ \alpha=1.5, 2.5 $ with an excitation at $ n_0=N/2 $. (a) Momentum excitation with $ B=2 $. In the inset, obtained energy density data of the system with $ \alpha=1.5 $ are fitted by $ |n-n_0|^{-5/3} $ in the asymptotic regime. (b) Displacement excitation with $ A=2 $. In the inset, obtained energy density data of the system with $ \alpha=1.5 $ are fitted by $ |n-n_0|^{-3} $ in the asymptotic regime. \label{energydensity}}
\end{figure}

\begin{figure}
\resizebox{0.9\columnwidth}{!}{
\begin{tabular}{cc} 
\includegraphics[width=\textwidth]{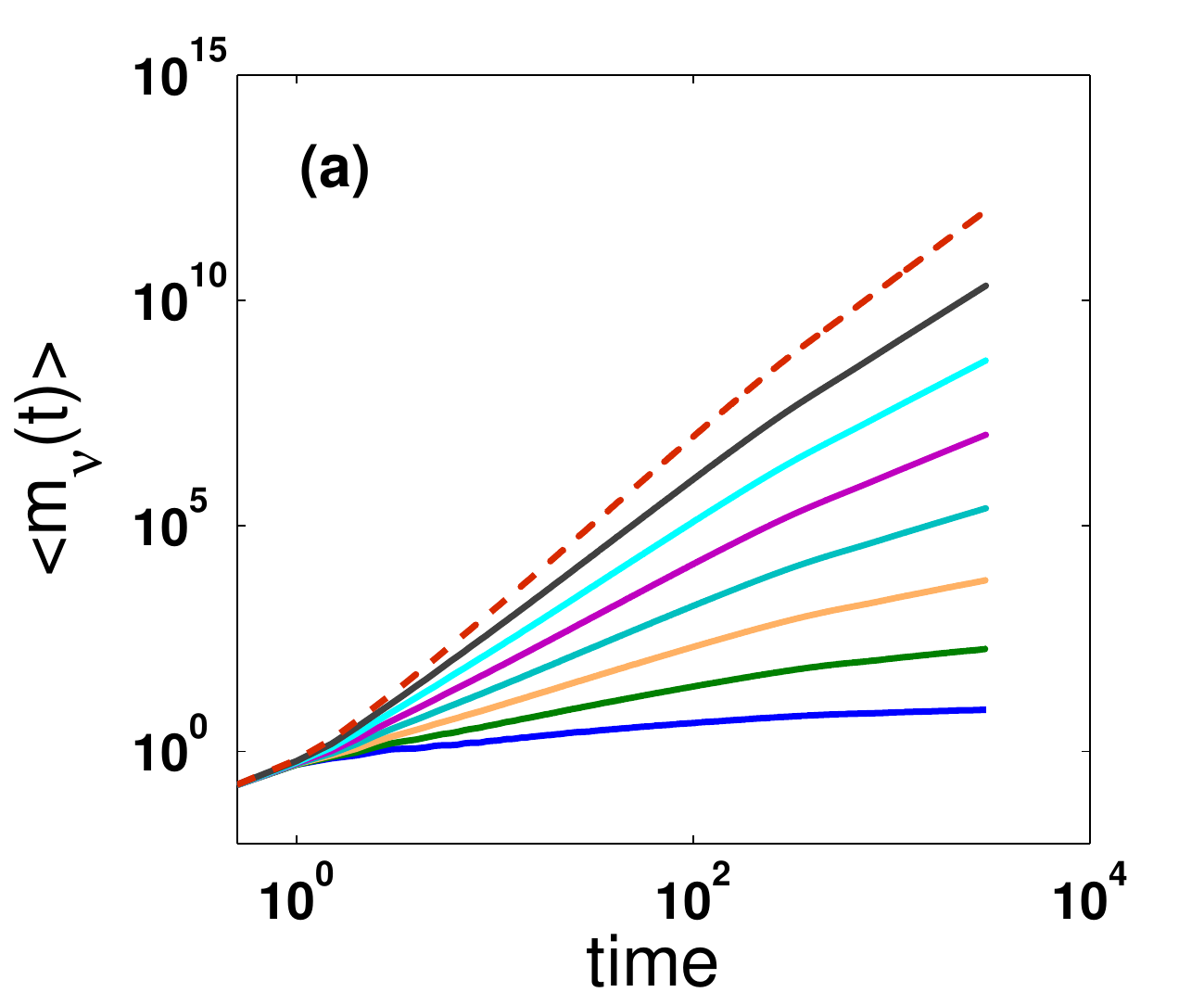} 
\includegraphics[width=\textwidth]{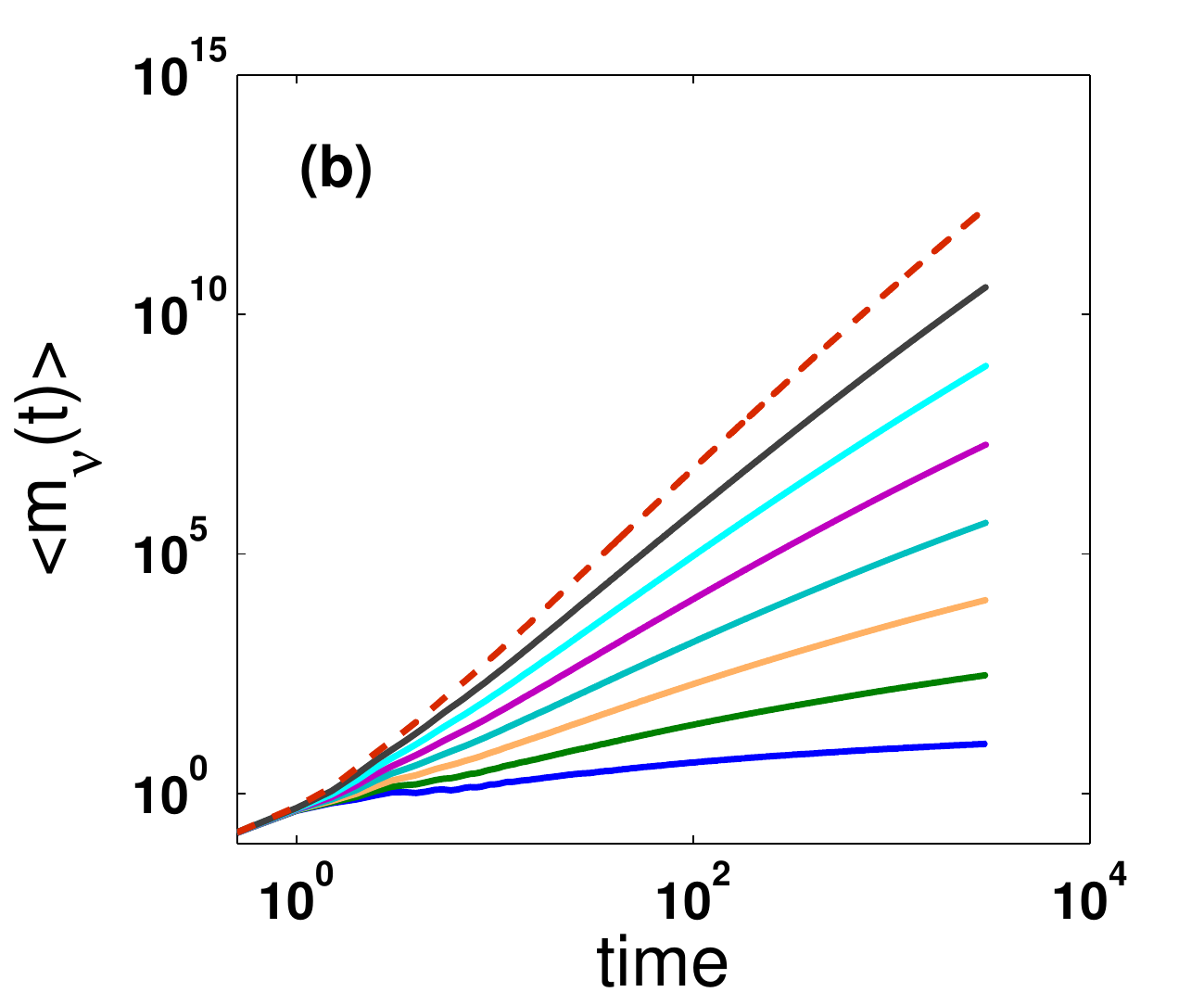} \\ 
\includegraphics[width=\textwidth]{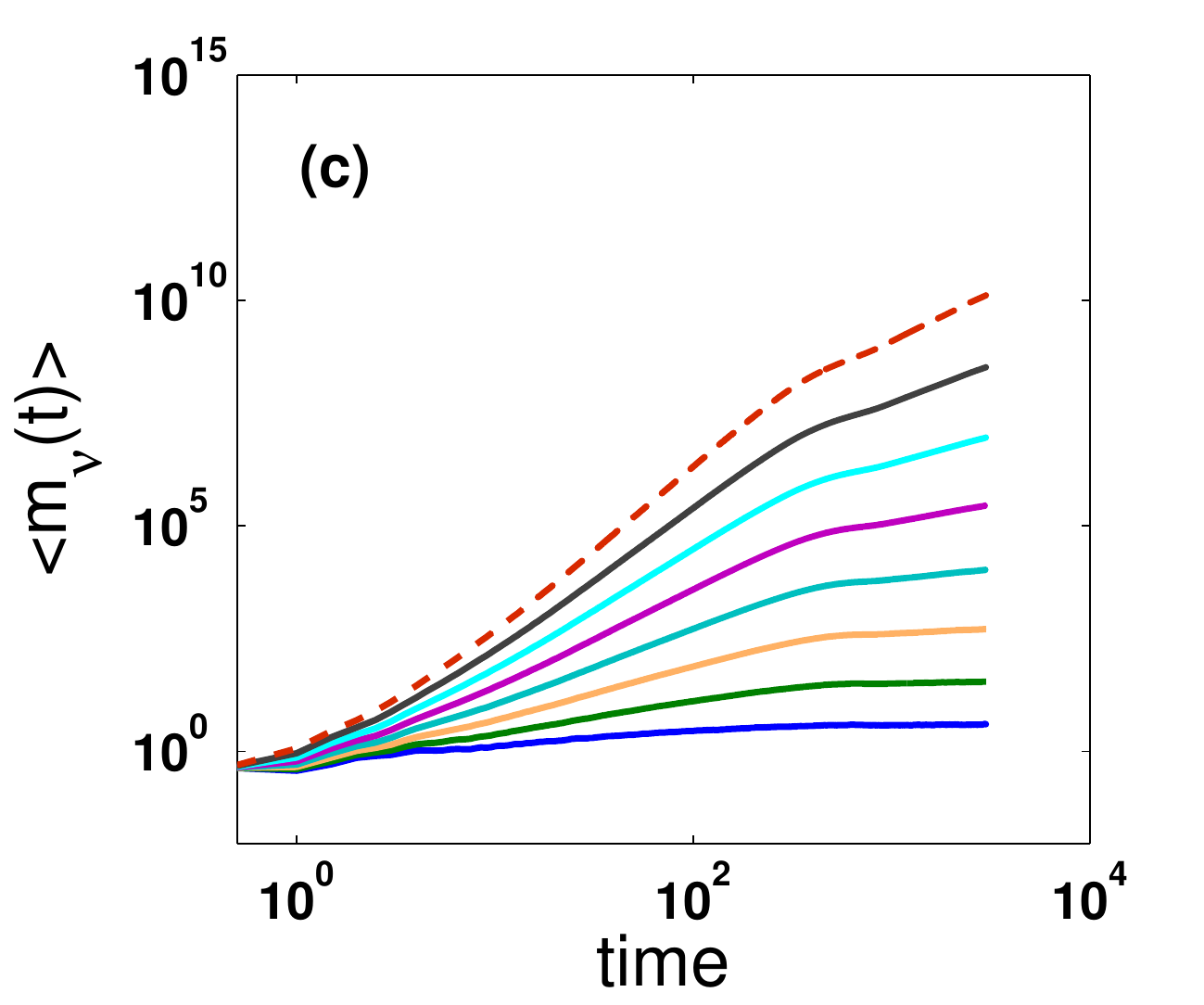} 
\includegraphics[width=\textwidth]{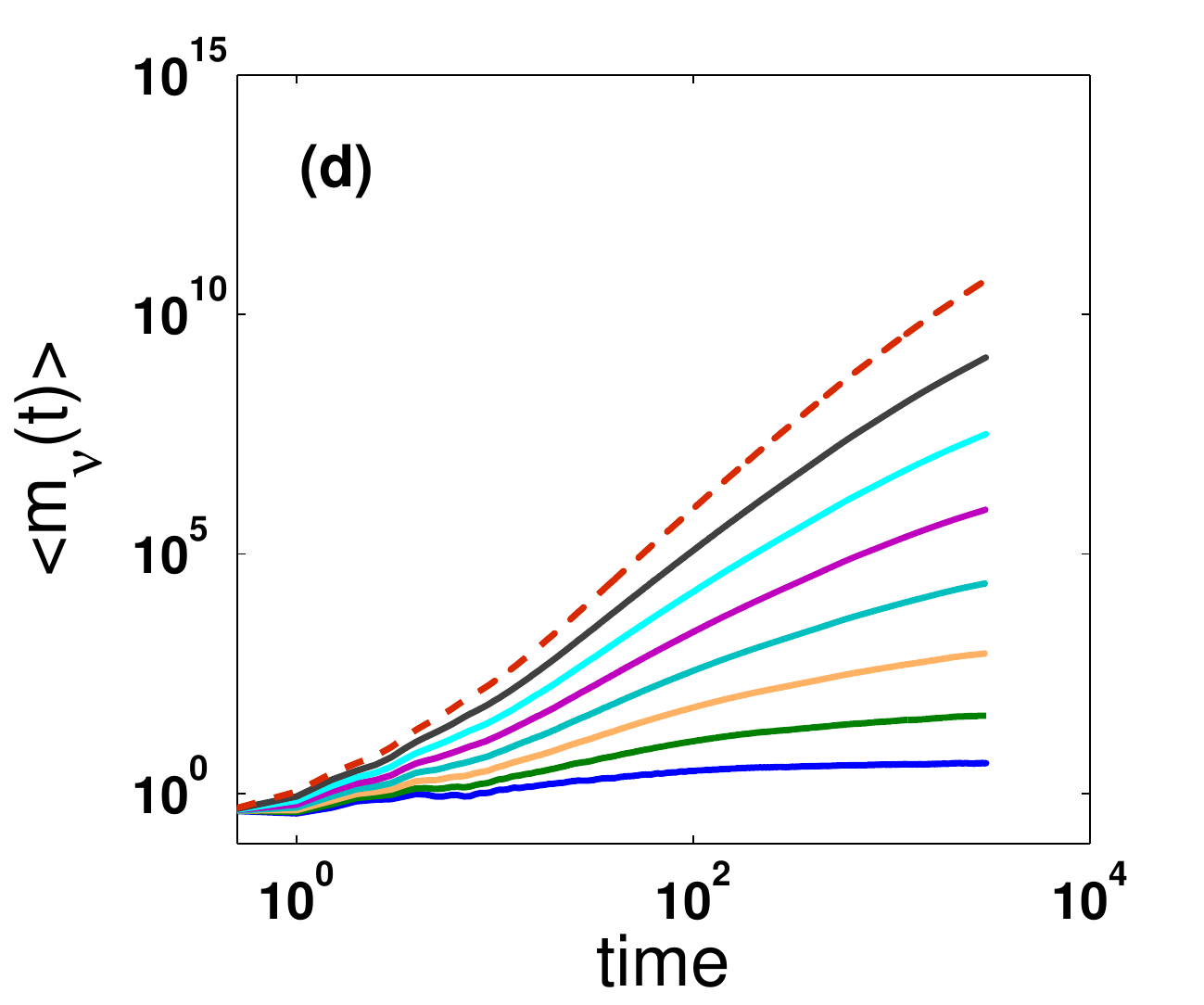} \\ 
\end{tabular}}

\caption{(color online) Time evolution of the moments $ \langle m_\nu (t) \rangle $ averaged over $ 2 \times 10^3 $ realizations for systems with length $ N=8 \times 10^3$. Different colours represent different index $ \nu $ varying from $ \nu=0.5 $ (solid blue) to $ \nu=4 $ (dashed red) with step $ \Delta \nu=0.5 $. Momentum excitation ($ B=2 $) was applied at $ n_0=N/2 $ on a system with L\'{e}vy exponent (a) $\alpha=1.5 $ (b) $\alpha=2.5 $. Displacement excitation ($ A=2 $) was applied at $ n_0=N/2 $ on a system with L\'{e}vy exponent (c) $\alpha=1.5 $ (d) $\alpha=2.5 $. \label{moments}}
\end{figure}

In our analysis, disordered lattices of length $ N=8\times 10^3 $ with an excitation at the center $ n_0=N/2 $ were studied. Two different types of excitation were applied: (i) displacement excitation with $ u_n(0)= A\delta_{n,n_0}$ and $ P_n(0)\equiv 0 $; (ii) momentum excitation with $P_n(0)=B\delta_{n,n_0}$ and $ u_n(0)\equiv 0 $. As it is known, 
these classes of initial condition yield different asymptotic properties 
due to the fact that the mode energy distribution is different in the two cases \cite{Datta1995}.
Based on the conservative dynamics of the system and by using a fourth-order symplectic algorithm \cite{McLachlan1992}, the set of coupled governing equations were numerically solved. We mainly focused on lattices with $ \alpha=1.5 $ and $ \alpha=2.5 $ to enable comparison of the two different regimes of transport. Note that the number of defects $ N_d $ on the considered lattices were different as $ N $ was kept fixed. Obtained results were then averaged over $ 2\times 10^3 $ realizations. Fig. \ref{energydensity} presents spread of the averaged energy density over the two systems after $ t=3 \times 10^3 $ for both  momentum and displacement excitations. In addition to a faster spread of energy, a broadening appeared around the excitation for the system with $ \alpha=1.5 $.

Following the analytical approach of ref.~\cite{Lepri2010} and assuming the frequency-dependent localization length as $ \xi_\alpha(\omega)\propto \omega^{-\alpha} $ in the small-frequency regime ($ \omega \rightarrow 0 $), the time and disorder averaged energy in a L\'{e}vy-type disordered system for a displacement excitation can be written as
\begin{equation}\label{Asymp-displacement}
\langle \overline{e_n(t)} \rangle \propto \int_{0}^{\infty} \omega^{2+\alpha}e^{-|n-n_0|\omega^\alpha}d\omega,
\end{equation}which asymptotically decays as $ |n-n_0|^{-(1+3/\alpha)} $. For a momentum excitation, $ \langle \overline{e_n(t)} \rangle $ can be similarly described by 
\begin{equation}\label{Asymp-momentum}
\langle \overline{e_n(t)} \rangle \propto \int_{0}^{\infty} \omega^{\alpha}e^{-|n-n_0|\omega^\alpha}d\omega,
\end{equation}with asymptotic behaviour as $ |n-n_0|^{-(1+1/\alpha)} $. The insets in Fig. \ref{energydensity} show the excellent agreement between our numerical results and the analytical predictions of the $ \alpha $-dependent power-law decay.  In the case of momentum excitation, data for $ \alpha=1.5 $ are fitted by $ |n-n_0|^{-5/3} $ with a goodness of $ 95 \% $ as shown in the inset in Fig. \ref{energydensity}(a). For the displacement excitation, data for $ \alpha=1.5 $ are fitted by $ |n-n_0|^{-3} $ with a goodness of $ 95 \% $ which is illustrated in Fig. \ref{energydensity}(b). \\

Moreover, another interesting entity is time evolution of the moments of the Hamiltonian which can be defined as \cite{Datta1995,Lepri2010}
\begin{equation}\label{moment definition}
m_\nu (t)=\dfrac{\sum_n \lvert n-n_0 \rvert^\nu e_n(t)}{H},
\end{equation} where $ \nu $ is a positive number and $ m_\nu (t) $ represents the $ \nu $th moment of the Hamiltonian at time $ t $. The second moment quantifies degree of the spreading of the wavepacket.

\begin{figure} 
\resizebox{0.99\columnwidth}{!}{
\begin{tabular}{cc} 
\includegraphics[width=\textwidth]{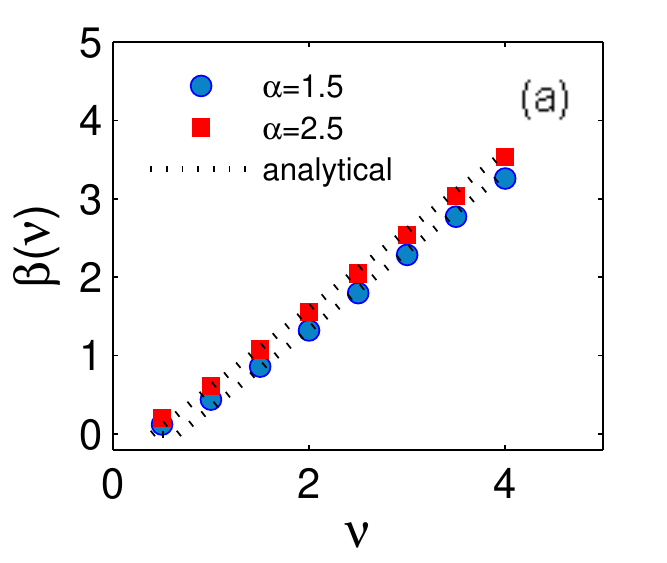} 
\includegraphics[width=\textwidth]{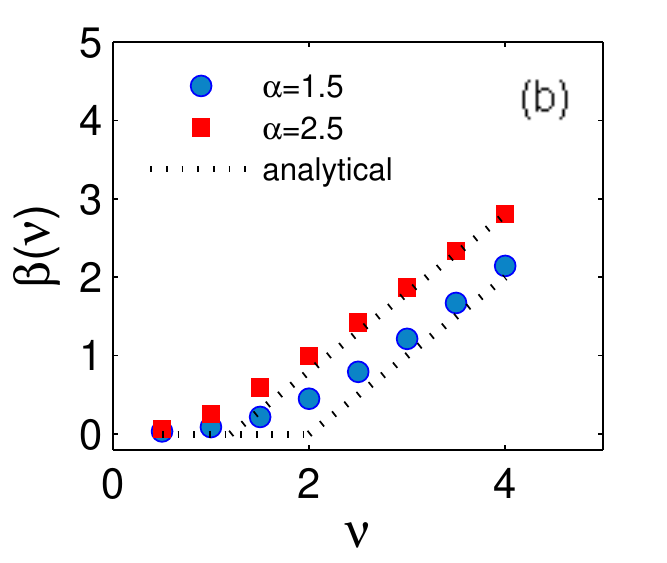} 
\end{tabular}}
\caption{(color online) Comparison of $ \beta(\nu) $ for the two systems with L\'{e}vy exponent $\alpha=1.5$ (blue circles) and $\alpha=2.5$ (red squares). Data are obtained by performing power-law fits on $ \langle m_\nu (t) \rangle $. Dotted black lines show the numerical solutions of analytical Eqs. (\ref{betanu-dis}) and (\ref{betanu-moment}) for each system. (a) Momentum excitation ($ B=2 $) (b) Displacement excitation ($ A=2 $). \label{beta}}
\end{figure}

In Fig. \ref{moments}, time evolution of different moments $ \langle m_\nu (t) \rangle $ of a wavepacket are shown and compared for the two systems. Beyond the clear signature of the growth of the disorder averaged moments over time, it was observed that around $ t\approx 10^3 $, the growth exponent suddenly changed in the system with $ \alpha=1.5 $. At shorter times, however, we may assume that correlation effects are not yet strong and system behaved like a short-range correlated one. Note that such effect is even more emphasized in the case of displacement excitation. To numerically estimate the scaling of  $ \langle m_\nu (t) \rangle \propto t^{\beta(\nu)}$, a power-law fit was performed on the averaged moments. The obtained results shown in Fig. \ref{beta}, reveal that $ \beta(\nu) $ is smaller in the system with $ \alpha=1.5 $ in comparison with  $ \alpha=2.5 $. \\

As already shown and discussed, $ \beta(\nu) $ is different in systems with different L\'{e}vy parameters. In an effort to determine a general analytic relation for $ \beta(\nu,\alpha) $ that describes dependence on both parameters $ \nu $ and $ \alpha $, the asymptotic forms of Eqs. (\ref{Asymp-displacement}) and (\ref{Asymp-momentum}) were inserted into Eq. (\ref{moment definition}). Later, by defining an upper cutoff for the summation in the numerator at the ballistic distance $ |n-n_0|=ct $, it can be shown that for displacement excitation 
\begin{equation}
\beta(\nu,\alpha) = \begin{cases} 
\nu-\frac{3}{\alpha}  &\mbox{, } \alpha \nu>3 \\
0 &\mbox{,  } \alpha \nu<3 \end{cases},
\label{betanu-dis}
\end{equation}
while for momentum excitation
\begin{equation}
\beta(\nu,\alpha) = \begin{cases} 
\nu-\frac{1}{\alpha}  &\mbox{, } \alpha \nu>1 \\
0 &\mbox{, }  \alpha \nu<1 \end{cases}.
\label{betanu-moment}
\end{equation}
As shown in Fig. \ref{beta}, there is an excellent agreement between the theoretical 
estimates, (according to Eqs. (\ref{betanu-dis}) and (\ref{betanu-moment})), and the numerical data obtained 
by power-law fits on $  \langle m_\nu (t) \rangle $ in the long-time regimes.

\section{Summary and conclusion} \label{conclusion}

In this paper, we have studied the localization properties of classical lattice 
waves in the presence of power-law correlated disorder. The model is mathematically 
simple and the choice of the disorder is motivated by the samples known 
as L\'evy glasses \cite{Barthelemy2008}. In the first part, we computed 
frequency-dependent localization length and analyzed how different characteristic 
parameters can affect it. Using theoretical arguments and numerics, we have shown 
that in the regime $ 1\leq \alpha <2 $, 
$ \gamma(\omega) \sim \omega^\alpha $ so that $\gamma$ mostly decreases as $ \alpha $ increases and this trend remains the same even in the regime $ \alpha>2 $. 
The physical interpretation of this result is that, for small frequencies and long wavelengths,
the waves see an effective disorder whose fluctuation is scale-dependent. This stems from the fact that the disorder is, by construction, scale-free. 

Large ordered regions can exist inside the system depending on the L\'{e}vy exponent $ \alpha $.  As expected, we showed that eigenfunctions of the system are localized where the density of disorder is high and extended in the large dilute spaces inside the system. Moreover, due to the superdiffusive nature of the transport in the range  $ 1\leq \alpha< 2 $, we showed that the spatial growth of the root-mean-squared phase $ \langle \phi_n^2 \rangle $ is greater in systems with smaller values of $ \alpha $. 

The anomalous localization properties reflects in the problem of wavepacket spreading. An initially localized perturbation attains at large times a characteristic 
power-law decay as in the uncorrelated case \cite{Lepri2010,Krapivsky2011}. However
in the present case, the exponent of the decay depends on the exponent $\alpha$ 
of the disorder distribution. This implies that the disorder averaged
momenta $\langle m_\nu (t) \rangle$ must diverge with time,
although the initial local energy excitation does not spread
completely. The predictions are clearly supported by the numerical results. 
As a matter of fact, it should be remarked that 
consideration of $m_2(t)$ alone is not sufficient to conclude that
the energy diffuses. Clearly, the origin of such a growth is drastically different 
to a genuinely (possibly anomalous) diffusive processes observed in nonlinear oscillator chains 
\cite{Zavt1993,Cipriani05,Delfini07a,Mulansky2010} and just stems from the slowly
decaying tails.

In the present work we mostly focused on the case $\alpha>1$. As 
argued in section \ref{tmat}, the case $\alpha<1$ should probably correspond
to a vanishing Lyapunov exponent. This may indicate that the localization 
in this regime is not standard. This is further supported by the similarity
of the problem with the perturbation growth in strongly intermittent dynamical
systems which indeed yields a vanishing Lyapunov exponent and 
a subexponential growth of perturbations \cite{Gaspard1988}. 
Moreover, the divergence of the variance of the distances should lead 
to huge sample-to-sample fluctuations. Some preliminary data confirm this
expectation suggesting that this case deserves a more detailed future study.

Finally, a natural question regards the extension to higher dimensions. 
Provided one is able to extend the definition of disorder, we expect the 
problem to be considerably more difficult in view of the nature 
of the predicted essential singularity of the localization length at vanishing frequencies 
\cite{John1983}.

\begin{acknowledgments}
Authors wish to thank Mario Mulansky for reading the manuscript and the financial support from the European Research Council (FP7/2007-2013), ERC Grant No. 291349.
\end{acknowledgments}

\bibliographystyle{apsrev}
\bibliography{bibnew}

\end{document}